\documentclass[conference]{IEEEtran}

\usepackage[T1]{fontenc}
\usepackage{graphicx}
\usepackage{cite}
\usepackage{caption}
\usepackage{subcaption}
\usepackage{epstopdf}

\usepackage{overpic}
\usepackage{amssymb}
\usepackage{amsmath}
\usepackage{amsthm}
\usepackage{array}
\usepackage{color}
\usepackage{url}

\IEEEoverridecommandlockouts

\theoremstyle{plain}

\newtheorem{lemma}{Lemma}

\newcommand{\vect}[1]{\mathbf{#1}}
\newcommand{\maximize}[1]{{\underset{{#1}}{\mathrm{maximize}}}}

\def\sinc{\mathrm{sinc}}
\def\Htran{\mbox{\tiny $\mathrm{H}$}}
\def\Ttran{\mbox{\tiny $\mathrm{T}$}}
\def\CN{\mathcal{N}_{\mathbb{C}}} 
\def\imagunit{\mathsf{j}} 

\begin{document}

\title{Utility-Based Precoding Optimization Framework for Large Intelligent Surfaces\vspace{-0.4cm}}

\author{
\IEEEauthorblockN{Emil Bj{\"o}rnson\IEEEauthorrefmark{1},
Luca Sanguinetti\IEEEauthorrefmark{2}
\thanks{E. Bj{\"o}rnson was supported by ELLIIT and CENIIT. L. Sanguinetti was supported by Pisa University under the PRA 2018-2019 
Project CONCEPT.}}
\IEEEauthorblockA{\IEEEauthorrefmark{1}\small{Department of Electrical Engineering (ISY), Link\"{o}ping University, Link\"{o}ping, Sweden (emil.bjornson@liu.se)}}
\IEEEauthorblockA{\IEEEauthorrefmark{2}\small{Dipartimento di Ingegneria dell'Informazione, University of Pisa, 56122 Pisa, Italy (luca.sanguinetti@unipi.it)}\vspace{-0.4cm}}
}

\maketitle

\begin{abstract}
The spectral efficiency of wireless networks can be made nearly infinitely large by deploying many antennas, but the deployment of very many antennas requires new topologies beyond the compact and discrete antenna arrays used by conventional base stations. In this paper, we consider the large intelligent surface scenario where small antennas are deployed on a large and dense two-dimensional grid. Building on the heritage of MIMO, we first analyze the beamwidth and sidelobes when transmitting from large intelligent surfaces. We compare different precoding schemes and determine how to optimize the transmit power with respect to different utility functions.
\end{abstract}


\vspace{-2mm}

\IEEEpeerreviewmaketitle

\section{Introduction}

A \emph{large intelligent surface} (LIS) consists of a physically large and dense antenna array \cite{Hu2018a}. Ideally, it is a continuous surface with controllable electromagnetic properties \cite{Hu2018a}, but we will consider also its discretized approximation. An LIS can be used for communication, positioning, and sensing \cite{Hu2018b}. If an LIS can be made thin, then it can potentially be integrated into walls.
It is well known that the beamwidth of the signal transmitted from an array is approximately inversely proportional to the aperture \cite{Tse2005a,massivemimobook}, thus there is a risk that two users have overlapping main beams also when using dense arrays \cite{Masouros2015a}.
In this paper, we investigate the shape of the sidelobes when transmitting from a continuous surface. In particular, we will explore if maximum ratio (MR) processing (as used in \cite{Hu2018a}) is sufficient when transmitting to multiple users or if more advanced methods, such as zero-forcing (ZF), is needed to deal with interference when transmitting from an LIS. It is known that MR is far from optimal in non-line-of-sight (NLOS) scenarios \cite{BHS18A}, but the situation might be different in line-of-sight (LOS) scenarios and when using dense arrays.

\section{Two Users and One LIS}

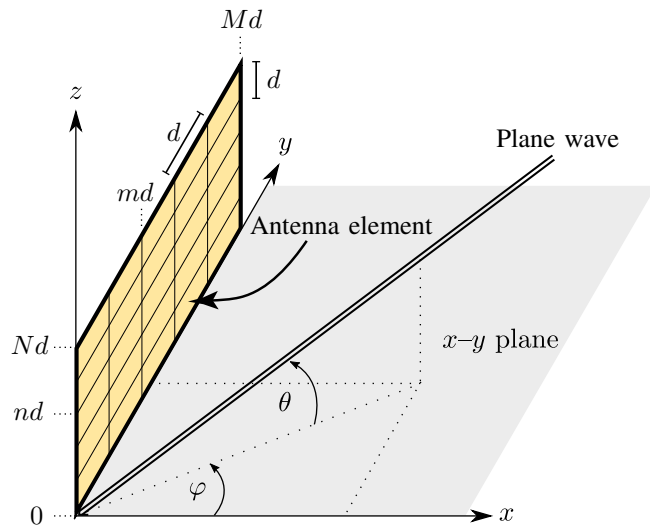
\begin{figure}[t!]
\begin{center}
	\begin{overpic}[width=\columnwidth,tics=10]{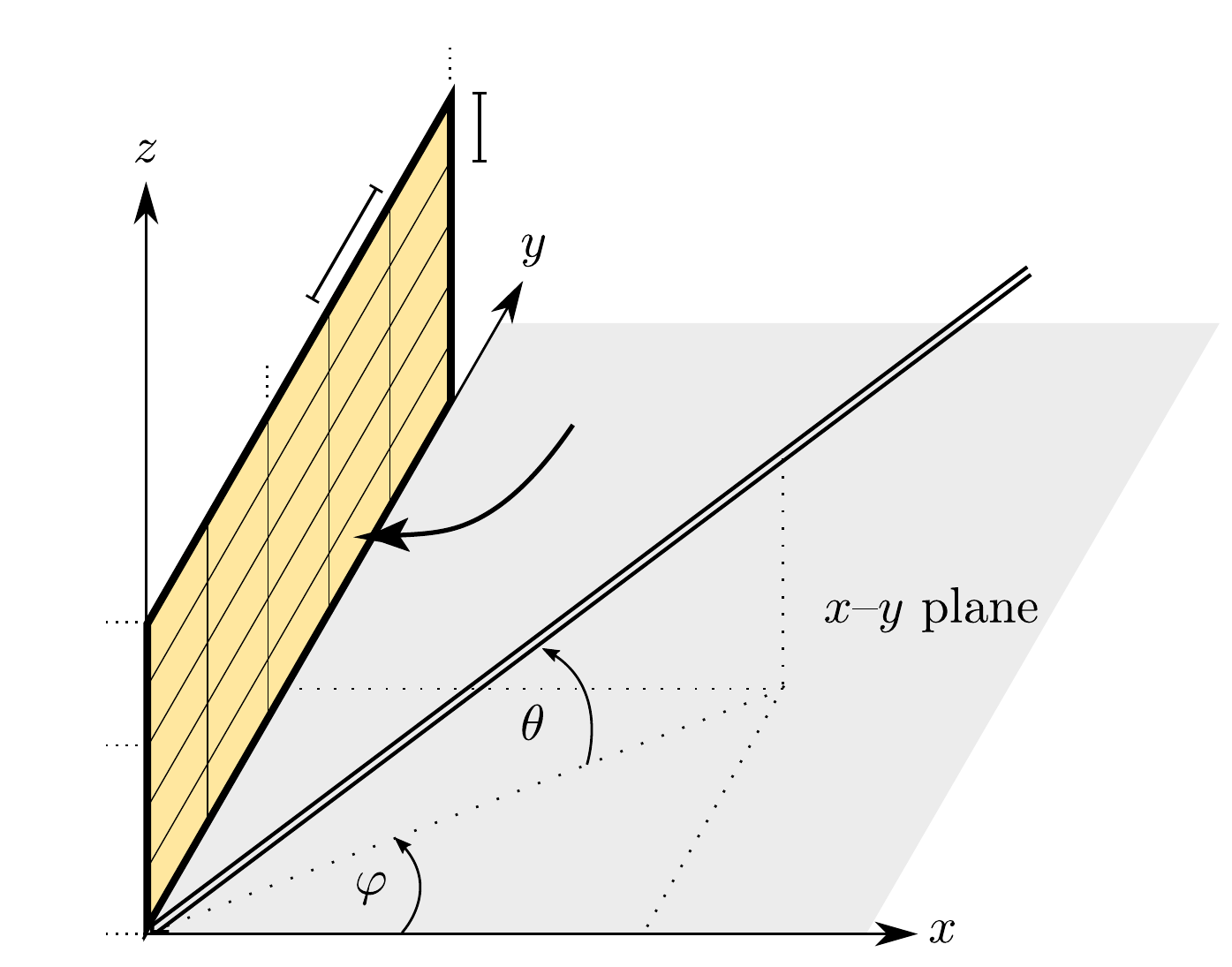}
		\put(5,1.5){$0$}
		\put(2.5,17){$nd$}
		\put(2.1,27){$Nd$}
		\put(18,50){$md$}
		\put(33.5,76){$Md$}
		\put(75,58){Plane wave}
		\put(38,45){Antenna element}
		\put(40.5,67){$d$}
		\put(25.5,59){$d$}
\end{overpic} 
\end{center} 
\caption{A planar array with $N$ rows and $M$ antennas per row. If a plane wave is impinging from elevation angle $\theta$ and azimuth angle $\varphi$, the array response vector is $\vect{a}(\varphi, \theta) $.\vspace{-0.3cm}} \label{figure_3DLoS}  
\end{figure}

To study the basic properties, we consider the scenario in which two users transmit to the large planar array, shown in Fig.~\ref{figure_3DLoS}. The array consists of $N$ horizontal rows and $M$ antennas per row, which are located in the $y$-$z$-plane. The array is dense in the sense that each element is of size $d \times d$, where $d$ is small, and the spacing is $d$ in the horizontal and vertical directions.\footnote{There will likely be mutual coupling between the antennas in a dense array, but this effect is neglected in this work to study the ideal situation. We refer to \cite{Hu2018c} for a study on mutual coupling in LIS.}

If a plane wave with wavelength $\lambda$ is impinging from  azimuth angle $\varphi \in [-\frac{\pi}{2},\frac{\pi}{2}]$ and elevation angle $\theta \in [-\frac{\pi}{2},\frac{\pi}{2}]$, we can compute the $NM \times 1$ array response vector using the methodology in \cite[Sec.~7.3.1]{massivemimobook}. If we order the antennas row by row, starting with the antenna in the origin, the array response vector can be expressed as 
\begin{equation}\label{eq:general-LoS-vector-channel}
\vect{a}(\varphi, \theta) = \left[e^{\imagunit  \alpha_{1,1}^{\varphi, \theta}} \, \ldots \, e^{\imagunit  \alpha_{1,M}^{\varphi, \theta}} \, e^{\imagunit  \alpha_{2,1}^{\varphi, \theta}} \, \ldots \, e^{\imagunit  \alpha_{N,M}^{\varphi, \theta}} \right]^{\Ttran}
\end{equation}
where  $\imagunit$ is the imaginary unit and the phase-shift of the antenna at row $n$ and column $m$ is
\begin{equation}
\alpha_{n,m}^{\varphi, \theta} = \frac{2\pi d}{\lambda} \big( (m-1) \cos (\theta) \sin(\varphi) + (n-1) \sin(\theta) \big).
\end{equation}
Note that $\| \vect{a}(\varphi, \theta) \|^2 = NM$. When transmitting from the array, $\vect{a}(\varphi, \theta)$ is the channel vector to a user located in the far-field in azimuth angle $\varphi$ and elevation angle $\theta$. The propagation distance should be larger than the Fraunhofer distance $\frac{2d^2 \max(M^2,N^2)}{\lambda}$ to use this far-field model.

We consider 
 two single-antenna users, which are located in the far-field of the array in the directions $(\varphi_1,\theta_1)$ and $(\varphi_2,\theta_2)$. We assume the angles are non-identical: either $\varphi_1 \neq \varphi_2$ and/or $\theta_1 \neq \theta_2$. The transmissions are precoded using the vectors $\sqrt{\rho_i} \vect{v}_i$ for user $i=1,2$. More precisely, $\vect{v}_i  \in \mathbb{C}^{NM}$ is the unit-norm precoding vector assigned to user $i$ and $\rho_i$ is the normalized transmit power, which represents the signal-to-noise ratio (SNR).
Moreover, let  $s_1,s_2$ denote the independent information-bearing signals with $\mathbb{E}\{ | s_1|^2 \} =  \mathbb{E}\{ | s_2|^2 \} = 1$.

The received signal at user 1 is modeled as
\begin{equation} \label{eq:received-2users}
y_1 = \vect{a}^{\Htran}(\varphi_1,\theta_1) \left(\sqrt{\rho_1} \vect{v}_1 s_1 + \sqrt{\rho_2}\vect{v}_2 s_2 \right)+ n_1
\end{equation}
where $n_1 \sim \CN(0, 1)$ is the normalized receiver noise. The received signal at user 2 is achieved by switching the user indices and is therefore omitted. By treating interference as noise, the achievable spectral efficiency (SE) is $\log_2(1+\mathrm{SINR}_1)$, where the signal-to-interference-and-noise ratio (SINR) is
\begin{equation} \label{eq:SINR1d}
\mathrm{SINR}_1 = \frac{ \rho_1 | \vect{a}^{\Htran}(\varphi_1,\theta_1) \vect{v}_1 |^2  }{ \rho_2 | \vect{a}^{\Htran}(\varphi_1,\theta_1) \vect{v}_2 |^2+1}.
\end{equation}
To further analyze this expression, we will now consider two different precoding schemes: MR and ZF precoding.

\vspace{-2mm}

\subsection{Maximum Ratio Transmission}

The first precoding scheme is MR 
\begin{equation}
\vect{v}_i = \frac{\vect{a}(\varphi_i,\theta_i)}{ \| \vect{a}(\varphi_i,\theta_i) \| }= \frac{\vect{a}(\varphi_i,\theta_i) }{ \sqrt{ NM} } \quad i=1,2
\end{equation}
which is optimal in noise-limited scenarios \cite{Bjornson2014d}.
When substituting these precoding vectors into \eqref{eq:SINR1d}, we obtain
\begin{equation} \label{eq:SINR1}
\mathrm{SINR}_1^{\mathrm{MR}} = \frac{\mathrm{SNR}_1}{{\mathrm{SNR}_2 \cdot\mathrm{I}_{12}^2} + 1}
\end{equation}
where  \vspace{-4mm}
\begin{align}\label{eq:interference}
\mathrm{I}_{12} & = \left| \frac{1}{NM}\vect{a}^{\Htran}(\varphi_1,\theta_1) \vect{a}(\varphi_2,\theta_2)\right|
\end{align}
accounts for the interference generated by user $2$ and $\mathrm{SNR}_i = NM\frac{\rho_i}{\sigma^2}$ for $i=1,2$ represents the received SNR of user $i$ in the absence of any interference. 
Observe that
\begin{align}\nonumber
\mathrm{I}_{12} &= \left|  \frac{1}{NM}\sum_{n=1}^{N} \sum_{m=1}^M e^{\imagunit (\alpha_{n,m}^{\varphi_2, \theta_2}  - \alpha_{n,m}^{\varphi_1, \theta_1} )} \right|\\
&=  \left| \frac{1}{N}\sum_{n=1}^{N} e^{\imagunit \frac{2\pi d}{\lambda} (n-1)\Omega} \right| \left| \frac{1}{M}\sum_{m=1}^{M} e^{\imagunit \frac{2\pi d}{\lambda} (m-1) \Psi} \right| \label{eq:signature_product}
\end{align}
where \vspace{-4mm}
\begin{align}\label{eq:omega}
\Omega &= \sin(\theta_2) - \sin(\theta_1), \\
\Psi &= \cos(\theta_2)\sin(\varphi_2) - \cos(\theta_1)\sin(\varphi_1).\label{eq:psi}
\end{align} 

\begin{lemma} \label{lemma:geoseries}
For any integer $N\geq 1$ and real-valued $A$,
\begin{equation} \label{eq:geoseries}
\sum_{n=1}^{N} e^{\imagunit 2\pi (n-1) A} \ = \begin{cases} \frac{\sin( \pi N A)}{\sin( \pi A) } \, e^{\imagunit \pi (N-1) A}, &A \neq 0, \\
 N, & A = 0.
 \end{cases}
 \end{equation}
\end{lemma}
By using the above lemma, we can rewrite \eqref{eq:signature_product} as\footnote{For brevity, we use this notation also for $\Omega=0$ and $\Psi=0$, bearing in mind the alternative expression in \eqref{eq:geoseries}.}
\begin{align}  \label{eq:signature_product2}
\mathrm{I}_{12}  =
\frac{1}{NM}\frac{\sin( \pi N d \Omega /\lambda)}{\sin( \pi d \Omega /\lambda) }\frac{\sin( \pi M d \Psi /\lambda)}{\sin( \pi d \Psi /\lambda) }.
\end{align}

\subsection{ZF Precoding}

ZF is the optimal precoding scheme in interference-limited scenarios \cite{Bjornson2014d} and there are two equivalent definitions: using a pseudo-inverse or an orthogonal projection matrix  \cite[Sec.~3.4.2]{Bjornson2013d}.
 \pagebreak
We consider the latter formulation for which
\begin{align}
\vect{v}_i &= \frac{\vect{w}_i}{\| \vect{w}_i \|} \quad \textrm{for } i=1,2 \textrm{ with}
\end{align} \vspace{-2mm}
\begin{align}
\vect{w}_1 &= \left(\vect{I}_{NM} - \frac{1}{NM} \vect{a}(\varphi_2,\theta_2)  \vect{a}^{\Htran}(\varphi_2,\theta_2)  \right)  \vect{a}(\varphi_1,\theta_1)\\
\vect{w}_2 &= \left(\vect{I}_{NM} - \frac{1}{NM} \vect{a}(\varphi_1,\theta_1)  \vect{a}^{\Htran}(\varphi_1,\theta_1)  \right)  \vect{a}(\varphi_2,\theta_2).
\end{align}
These vectors satisfy the ZF properties $\vect{a}^{\Htran}(\varphi_2,\theta_2) \vect{v}_1 = 0$ and  $\vect{a}^{\Htran}(\varphi_1,\theta_1) \vect{v}_2 = 0$.
Hence, substituting $\vect{v}_i$ into \eqref{eq:SINR1d} yields\begin{align}
& \mathrm{SINR}_1^{\mathrm{ZF}} 
=\rho_1 \frac{\left| \vect{a}^{\Htran}(\varphi_1,\theta_1) \vect{w}_1 \right|^2}{\left\| \vect{w}_1  \right\|^2} 
= \mathrm{SNR}_1 (1  - \mathrm{I}_{12}^2) .
\label{eq:SINR1ZF}
\end{align}
Interestingly, \eqref{eq:SINR1ZF} contains the same components as \eqref{eq:SINR1} ($ \mathrm{SNR}_1$, $\mathrm{I}_{12}^2$, and $1$), but has a different structure. 
In \eqref{eq:SINR1ZF}, $\mathrm{I}_{12}^2$ should be interpreted as the perfomance loss due to the cancellation of the interference generated by user 2. 

\section{System Analysis for Dense Arrays}
We will now analyze the system above in the limit of infinitesimal antennas for a given array dimension, which represents an ideal LIS. More precisely, we fix the array's horizontal length to $L = Md$ and the vertical height to $H = Nd$, and then we will let $d \to 0$.
Each antenna has a physical size of $d \times d$ but the effective size shrinks to $d \cos(\varphi_i) \times d\cos(\theta_i)$ when observing it from the direction of user $i$. The SNR per antenna reduces with the effective antenna area \cite{friis1946note}, which in free-space propagation is modeled as
\begin{align} \notag
\frac{\mathrm{SNR}_i}{NM} &= \frac{q_i}{\sigma^2}\underbrace{d^2\cos(\varphi_i) \cos(\theta_i)}_{\text{Effective area}} \underbrace{\frac{1}{ 4\pi r_i^2}}_{\text{Free-space propagation}} \\&= d^2 \hspace{-0.7cm}\underbrace{\frac{q_i}{\sigma^2}\frac{ \cos(\varphi_i) \cos(\theta_i)}{ 4\pi r_i^2}}_{= p_i\text{,\;Independent of the antenna size $d$}} \label{eq:rho_i}
\end{align}
for $i=1,2$,
where $q_i$ is the unnormalized transmit power, $\sigma^2$ is the noise power, $r_i$ is the distance to user $i$. As indicated in \eqref{eq:rho_i}, we denote the part that does not depend on $d$ as $p_i$.

\subsection{Limiting SINRs}

By using \eqref{eq:rho_i}, we have that
\begin{equation} \label{eq:signal-dense-limit}
\mathrm{SNR}_i = NM\frac{\rho_1}{\sigma^2} = p_i d^2 NM = p_i LH \quad \textrm{for} \,\,i=1,2,
\end{equation}
 depends on the array area $LH$ but not on the area $d^2$ of each antenna. However, the interference gain in  \eqref{eq:signature_product2} depends on $d$:
\begin{align}\label{eq:interference_fixed_array}
\mathrm{I}_{12}^2&=
\frac{1}{(NM)^2} 
{\frac{\sin^2( \pi H \Omega/\lambda)}{\sin^2( \pi d \Omega/\lambda) } \frac{\sin^2( \pi L \Psi/\lambda)}{\sin^2( \pi d \Psi/\lambda) }}.
\end{align}
By letting $d \to 0$ and utilizing that $\sin(x) \approx x$ is a tight approximation as $x \to 0$, the following limit is obtained.

\begin{lemma}\label{lemma:limiting_interference}
If $d \to 0$, then
\begin{align}
\mathrm{I}_{12,d=0}^2=\lim_{d\to 0} \mathrm{I}_{12}^2  
= \sinc^2 \left( \frac{ H \Omega}{\lambda} \right) \sinc^2 \left( \frac{ L \Psi}{\lambda} \right) \label{eq:interference-dense-limit}
\end{align}
where $\sinc(x) = \sin(\pi x)/(\pi x)$ is the sinc-function. 
\end{lemma}
The above limit is in general non-zero, which was expected since the spatial resolution of an array is known to depend on the aperture (i.e., length $L$ and height $H$) and not the antenna spacing; see for example, \cite[Sec.~7.2.4]{Tse2005a}, \cite[Sec.~7.4.2]{massivemimobook}, \cite{Masouros2015a}. The two squared sinc-functions determine how large the interference is. Since the two users are assumed to have non-identical angles, we have $\Omega \neq 0$ and/or $\Psi \neq 0$, which implies that at least one of the sinc-functions can be small if the array is physically large.
By using Lemma~\ref{lemma:limiting_interference}, the limiting SINRs with MR and ZF easily follow.

\begin{lemma}
The limiting SINRs with MR and ZF are:
\begin{align} \notag
 \mathrm{SINR}_{1,d=0}^{\mathrm{MR}} &= \lim_{d\to 0} \mathrm{SINR}_1^{\mathrm{MR}} \\ &= \frac{  p_1 LH  }{
p_2 LH \sinc^2 \left( \frac{ H \Omega}{\lambda} \right) \sinc^2 \left( \frac{ L \Psi}{\lambda} \right)+ 1} \label{eq:SINR1limit}
\end{align}
\begin{align} \notag
 \mathrm{SINR}_{1,d=0}^{\mathrm{ZF}} &=
\lim_{d\to 0} \mathrm{SINR}_1^{\mathrm{ZF}} \\ &= p_1 LH \left( 1 - \sinc^2 \left( \frac{  H \Omega}{\lambda} \right) \sinc^2 \left( \frac{ L \Psi}{\lambda} \right) \right).  \label{eq:SINR1limitZF}
\end{align}
\end{lemma}

\subsection{Interference Gain: Beamwidth and Sidelobes}

We now analyze the interference gain when user~1 has $\varphi_1=\theta_1=0$, while the interfering user 2 has $\theta_2=0$ but a varying azimuth angle $\varphi_2 \in [-\pi/2,\pi/2]$. From \eqref{eq:omega} and \eqref{eq:psi}, we thus obtain $\Omega = 0$ and $\Psi=\sin(\varphi_2)$ such that \eqref{eq:interference_fixed_array} reduces to
\begin{align}\label{eq:interference_case_study}
\mathrm{I}_{12}^2&=
\frac{d^2}{L^2} 
{ \frac{\sin^2( \pi L \Psi/\lambda)}{\sin^2( \pi d \Psi/\lambda) }}
\end{align}
which shows that, for any given $\Psi$ and $L/\lambda$, the interference gain depends on $d$, where a smaller $d$ leads to smaller values. From Lemma~\ref{lemma:limiting_interference}, the limit is given by
\begin{align}\label{eq:limiting_interference_case_study}
\mathrm{I}_{12,d=0}^2 
= \sinc^2 \left( \frac{ L \Psi}{\lambda} \right).
\end{align}
The maximum of both \eqref{eq:interference_case_study} and \eqref{eq:limiting_interference_case_study} is achieved for $\varphi_2 = 0$, which makes $\Psi = 0$.
To find the nulls of the interference gain, \eqref{eq:interference_case_study} and \eqref{eq:limiting_interference_case_study} are set equal to zero, which leads to
\begin{align}
\varphi_2=\varphi_{2,n}^{\rm{null}} = \pm \arcsin \left(\frac{\lambda}{L} {n} \right)
\approx \pm \frac{\lambda}{L} {n} \quad n=1,2,\ldots
\end{align}
where the approximation holds for $L \gg \lambda n$.
If we define the beamwidth as the angular distance between the first two nulls, it is  approximately $2\lambda/L$. In line with classical results on the resolution of arrays \cite[Sec.~7.2.4]{Tse2005a}, the beamwidth is inversely proportional to the array length $L$, but independent of $d$. 

\begin{figure} 
        \centering\vspace{-2mm}
        \begin{subfigure}[t]{\columnwidth} \centering 
	\begin{overpic}[width=1.1\columnwidth,tics=10]{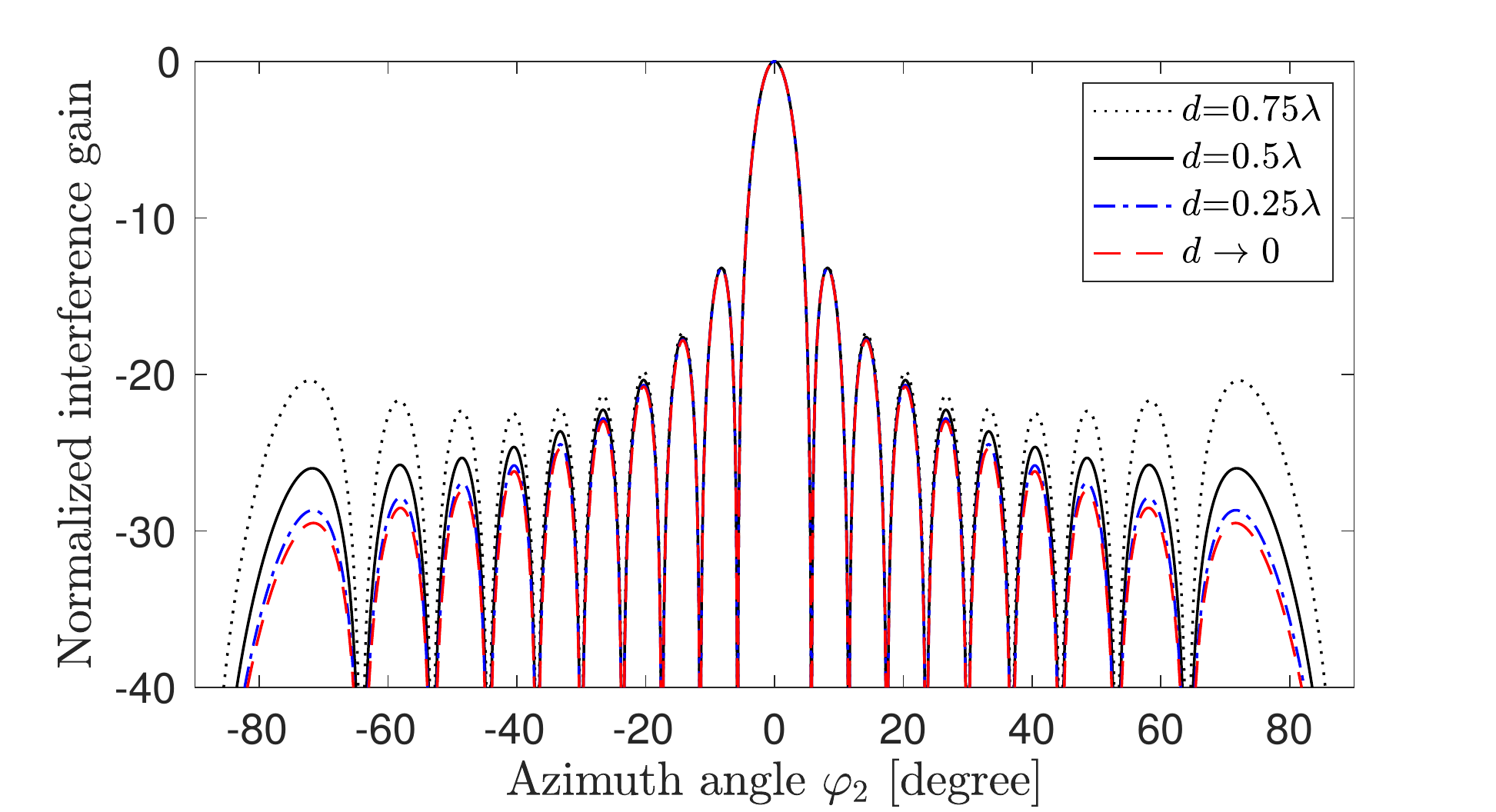}
\end{overpic} 
                \caption{$L=H=10\lambda$} 
                \label{simulationBeamPattern_1m} 
        \end{subfigure} 
        \begin{subfigure}[t]{\columnwidth} \centering  
	\begin{overpic}[width=1.1\columnwidth,tics=10]{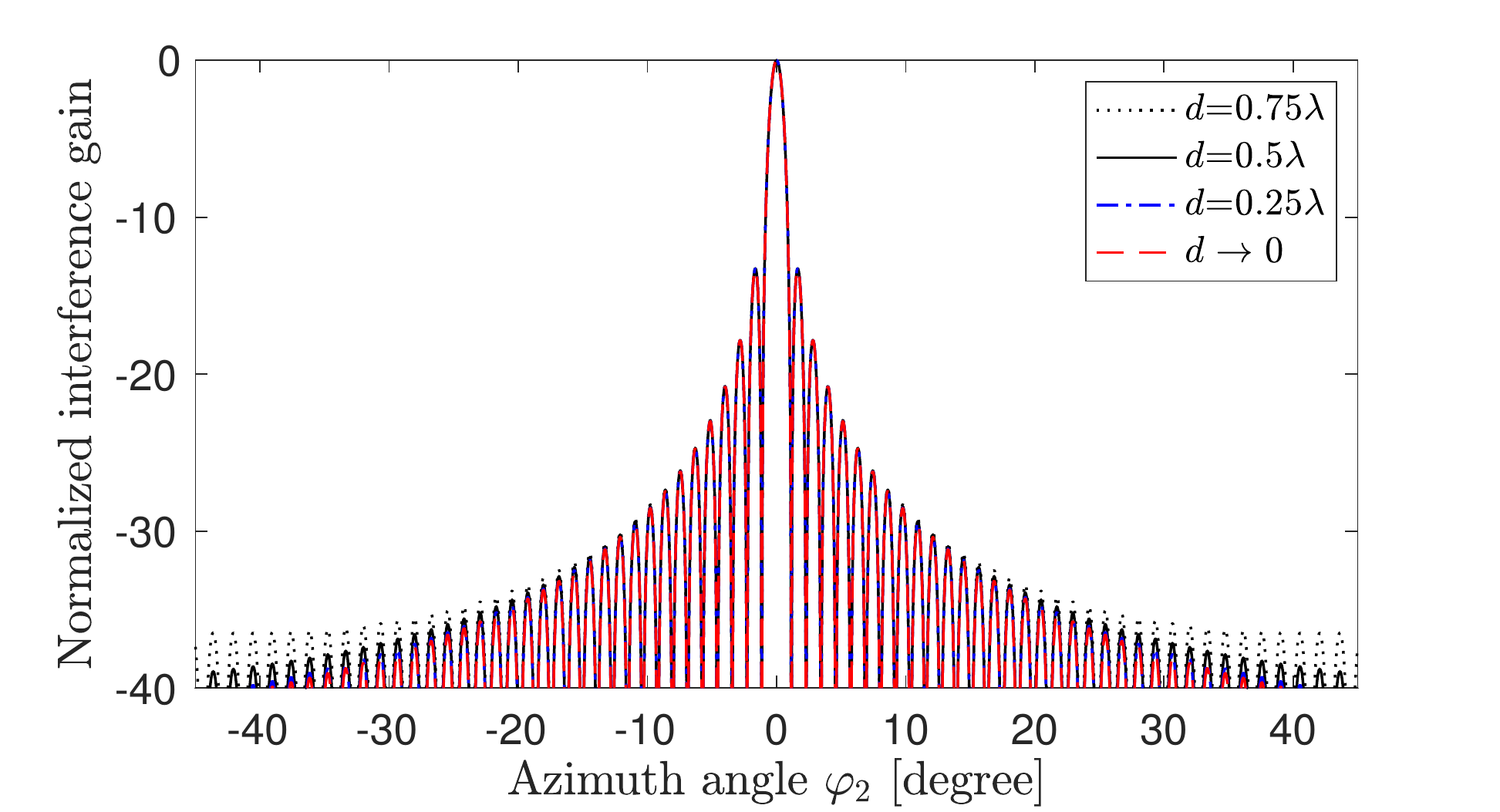}
\end{overpic} 
                \caption{$L=H=50\lambda$} 
                \label{simulationBeamPattern_5m}
        \end{subfigure} 
        \caption{The interference gain from a user located at $\theta_2=0$ and $\varphi_2 \in [-\pi/2,\pi/2]$, while the desired user is at $\varphi_1=\theta_1=0$.}
        \label{fig:simulationBeamPattern}   \vspace{-4mm}
\end{figure}

The maxima of the sidelobes occur when the numerator of \eqref{eq:interference_case_study} and \eqref{eq:limiting_interference_case_study} attains its maximum; that is, $\sin( \pi L \Psi/\lambda) = \pm 1$. For $L\gg \lambda n$, it is \emph{approximately} given by
\begin{align}\label{eq:maxima_of_sidelobes}
\varphi_2=\varphi_{2,n}^{\max} \approx \pm \frac{\lambda}{L} \frac{2n +1}{2} \quad n=1,2,\ldots.
\end{align}
By using the above approximation in \eqref{eq:interference_case_study} and \eqref{eq:limiting_interference_case_study}, we obtain
\begin{align}\label{eq:interference_case_study_value_n}
\left.\mathrm{I}_{12}^2\right|_{\varphi_{2}=\varphi_{2,n}^{\max}} \approx 
\frac{d^2}{L^2} 
{ \frac{\sin^2( \pi L \varphi_{2,n}^{\max}/\lambda)}{\sin^2( \pi d \varphi_{2,n}^{\max}/\lambda) }} = \frac{d^2}{L^2} 
{ \frac{1}{\sin^2\left( \frac{2n +1}{2} \pi d /L\right) }}
\end{align}
and
\begin{align}\label{eq:limiting_interference_case_study_value_n}
\left.\mathrm{I}_{12,d=0}^2\right|_{\varphi_{2}=\varphi_{2,n}^{\max}}\approx \left(\frac{2}{2n +1} \frac{1}{\pi}\right)^2
\end{align}
from which it follows that the maximum of the first sidelobe (i.e., $n=1$) of an ideal LIS is $(\frac{2}{3\pi})^2=-13.46$~dB weaker than the main lobe, irrespective of the surface length $L$. The difference between \eqref{eq:interference_case_study_value_n} and \eqref{eq:limiting_interference_case_study_value_n} depends on $d/\lambda$ and $n$, where smaller $d/\lambda$ and/or $n$ lead to smaller differences. As a rule-of-thumb, we can approximate \eqref{eq:interference_case_study_value_n} with \eqref{eq:limiting_interference_case_study_value_n} whenever $\pi d\varphi_{2,n}^{\max}/\lambda \le \pi^2/8$ since then $\sin(x) \approx x$ with an error below $10\%$. For $\varphi_{2,n}^{\max} \in [-\pi/4,\pi/4]$, we obtain $d\le \lambda/2$.

Fig.~\ref{fig:simulationBeamPattern} shows the interference gain for the antenna sizes $d\in \{0.25\lambda, \, 0.5\lambda, \, 0.75\lambda \}$ and as $d\to 0$.
Particularly, Fig.~\ref{fig:simulationBeamPattern}(a) considers a surface with size $L=H=10 \lambda$, while Fig.~\ref{fig:simulationBeamPattern}(b) considers $L=H=50 \lambda$. 
As expected, the spatially undersampled case of $d=0.75\lambda $ leads to the largest sidelobes. Interestingly, there is little difference between $d=0.25\lambda $ and $d \to 0$, which demonstrates that a discretization of the ideal continuous LIS concept will likely perform very well.

\subsection{MR versus ZF Precoding}

We will continue evaluating the SE that is achieved by user 1 with MR and ZF. We use the asymptotic expressions in \eqref{eq:SINR1limit} and \eqref{eq:SINR1limitZF} with $L=H= 50\lambda$.
User~1 is located at the angles $\varphi_1=\theta_1=0$, while we vary both angles for user 2. We assume the users are equipped with lossless omnidirectional antennas and set the SNRs to $\rho_1=\rho_2 = 20$\,dB. 

\begin{figure} 
        \centering
        \begin{subfigure}[t]{\columnwidth} \centering 
	\begin{overpic}[width=.95\columnwidth,tics=10]{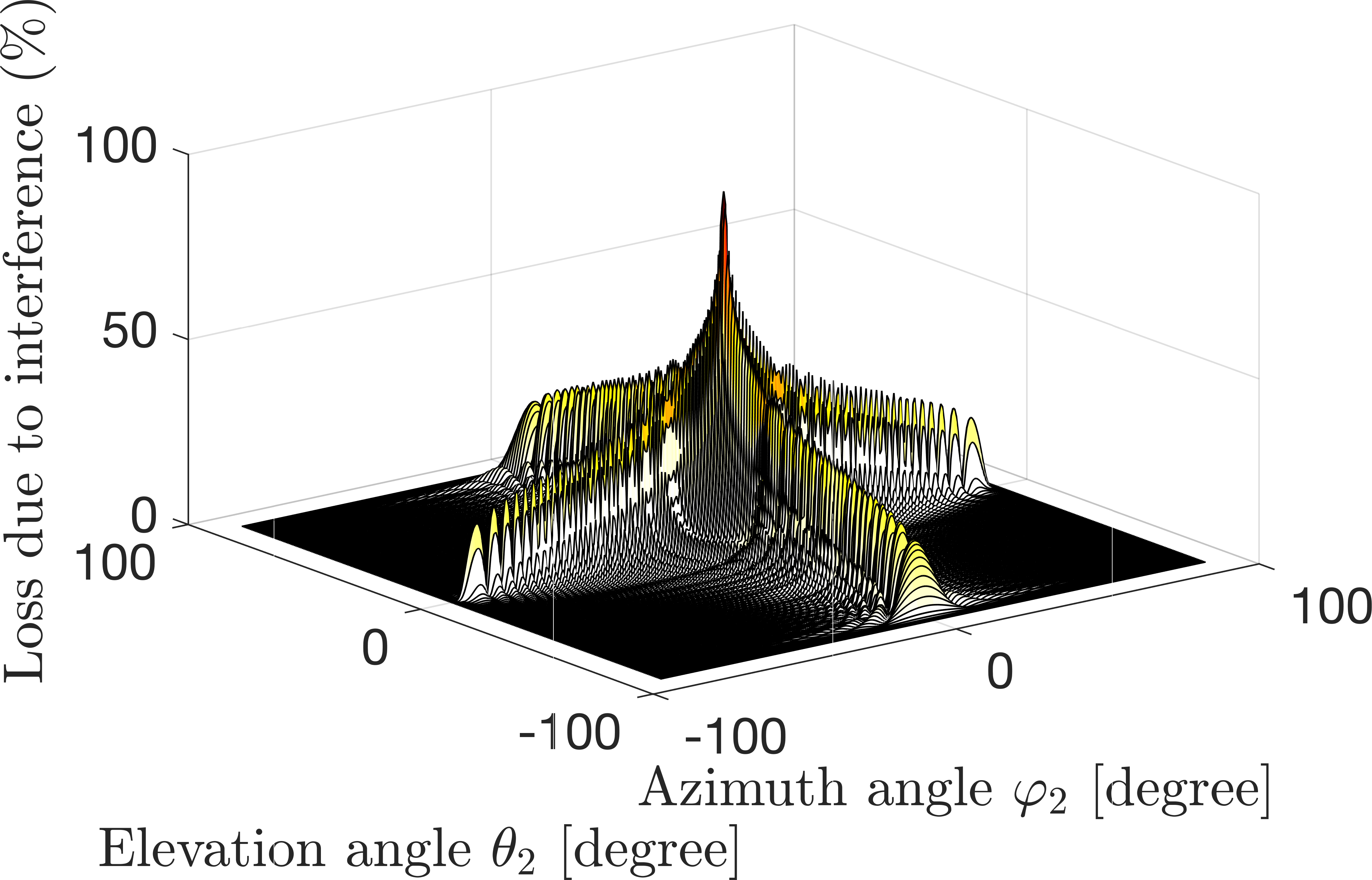}
\end{overpic} 
                \caption{MR precoding} 
                \label{simulation_downlink_angles}
        \end{subfigure} 
        \begin{subfigure}[t]{\columnwidth} \centering  
	\begin{overpic}[width=.95\columnwidth,tics=10]{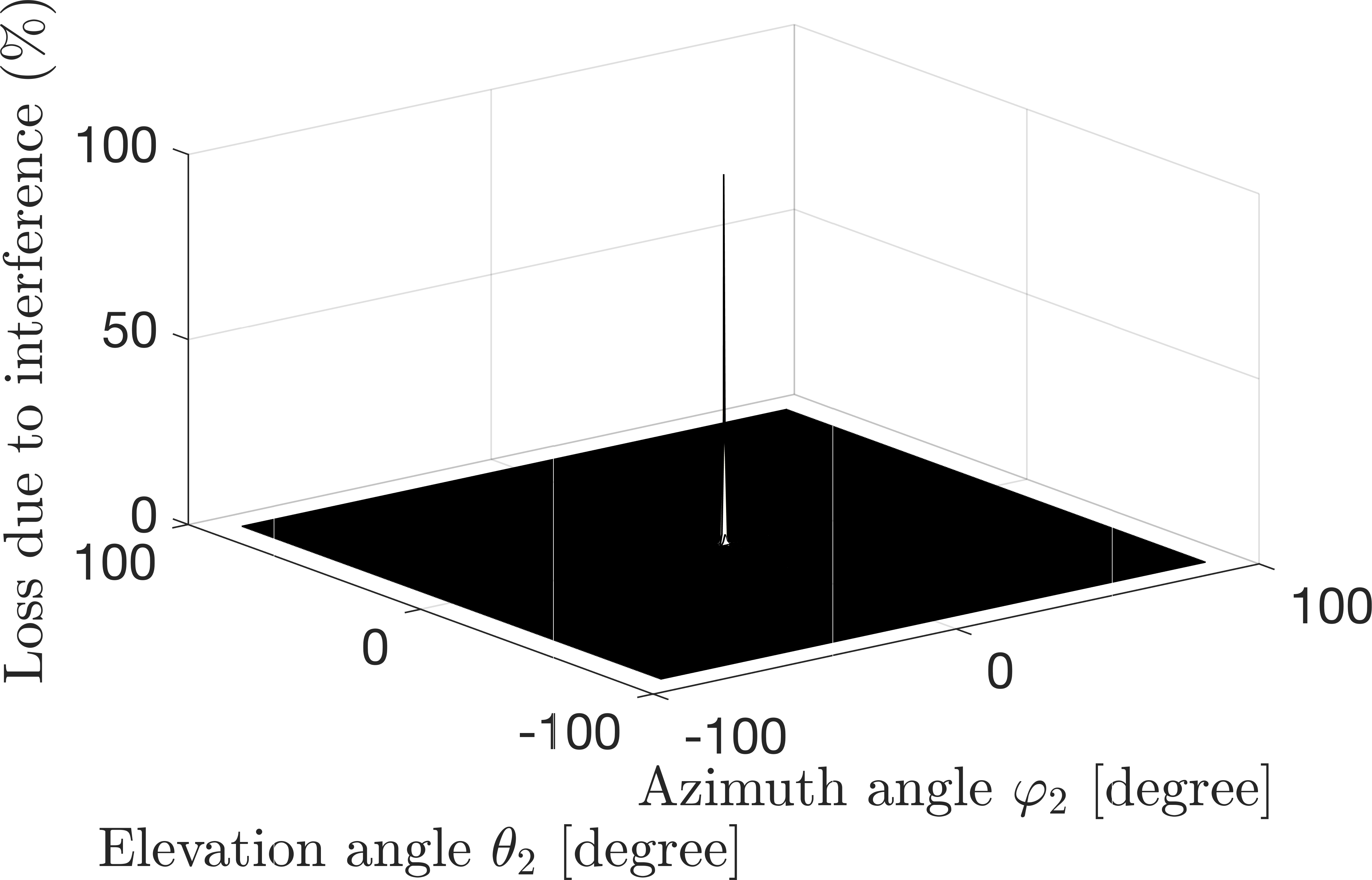}
\end{overpic} 
                \caption{ZF precoding} 
                \label{simulation_downlink_angles_ZF}
        \end{subfigure} 
        \caption{Relative loss in SE due to interference in a setup where the desired user is at $\varphi_1=\theta_1=0$ and the interfering user's angles are varied.}
        \label{fig:simulation_downlink_angles_MRT_ZF}   \vspace{-4mm}
\end{figure}

Fig.~\ref{fig:simulation_downlink_angles_MRT_ZF} shows the performance of MR and ZF in terms of the relative performance loss compared to an ideal interference-free case with SE $\log_2(1+\mathrm{SNR}_1)$. There are many angles $(\varphi_2,\theta_2)$ in Fig.~\ref{fig:simulation_downlink_angles_MRT_ZF}(a) where substantial interference is caused by user 2, so that MR suffers from reduced performance. This is particularly the case when either the azimuth angle or the elevation angle is similar to that of user 1. In contrast, the performance loss in Fig.~\ref{fig:simulation_downlink_angles_MRT_ZF}(b)  when using ZF is almost zero, except when the two users have identical angles---in that case, MR loses 95\% and ZF\% loses 100\% of the performance due to interference. The conclusion is that interference can be suppressed almost for free when using an LIS, but the active interference-suppression of ZF is needed---MR is greatly suboptimal just as in Massive MIMO with NLOS channels~\cite{BHS18A}.

\subsection{Asymptotic Analysis with $L \to \infty$}

Although ZF outperforms MR for a finite surface, the situation might change as the surface grows.
We can let the surface grow large, for example, by letting $L \to \infty$. This limit is not practically achievable using our far-field channel model, but it is still accurate for very large arrays \cite{Bjornson2019e}.

\begin{figure}[t!]
\begin{center}
	\begin{overpic}[width=\columnwidth,tics=10]{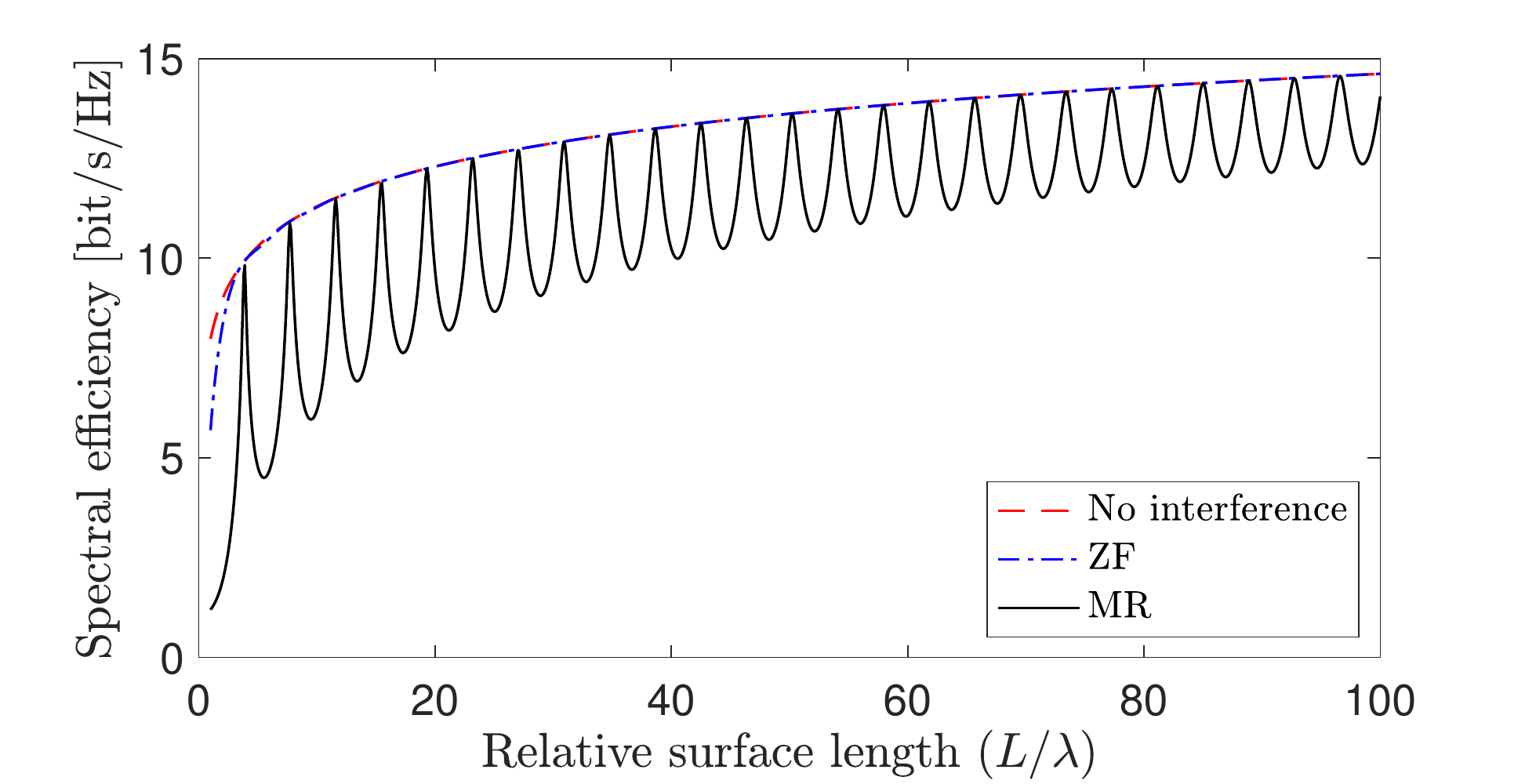}
\end{overpic} 
\end{center} \vspace{-2mm}
\caption{SE behavior with ZF and MR as $L$ increases.} \label{simulation_downlink_difference} \vspace{-4mm}
\end{figure}

Since $L \sinc^2 ( \frac{ \pi L \Psi}{\lambda}) \leq \frac{\lambda^2}{ \pi^2 L \Psi^2}$ for $\Psi\neq 0$, where the upper bound goes to zero, the interference/loss terms containing $\mathrm{I}_{12}^2$ in \eqref{eq:SINR1limit} and \eqref{eq:SINR1limitZF} go asymptotically to zero and we get:
\begin{lemma}
In the regime of an asymptotically wide LIS, 
\begin{equation}
\lim_{L\to \infty} \log_2 \left( 1 + \mathrm{SINR}_{1,d=0}^{\mathrm{ZF}} \right) - \log_2\left( 1 + \mathrm{SINR}_{1,d=0}^{\mathrm{MR}} \right) = 0.
\end{equation}
\end{lemma}
Hence, the performance difference between ZF and MR disappears as the LIS size increases. This is different from the classical i.i.d.~Rayleigh fading case where there is always a performance gap \cite{Rusek2013a}.
Fig.~\ref{simulation_downlink_difference} shows the SE as a function of $L$ when using ZF, MR, and when there is no interference. We assume $H= \lambda$, $\varphi_1=\theta_1=\theta_2=0$, $\varphi_2=\pi/12$, and a reference SNR of $\rho_1=\rho_2 = 20$\,dB. ZF quickly converges to the interference-free case, while MR gives oscillations since the increasing surface moves the locations of the sidelobes.

\section{Precoding Optimization Framework}

We now consider a more general setup where the LIS serves $K$ users, where user $k$ has SNR $\rho_k$ and is located in direction $(\varphi_k,\theta_k)$. The users have non-identical angle pairs (i.e., $\varphi_k \neq \varphi_i$ and/or $\theta_k \neq \theta_i$ for $k \neq i$). The LIS transmits to user $i$ using the normalized transmit power $\rho_i$ and unit-norm precoding vector $\vect{v}_i$. The received signal at user $k$ is
\begin{equation} \label{eq:received-Kusers}
y_k = \vect{a}^{\Htran}(\varphi_k,\theta_k) \sum_{i=1}^{K} \sqrt{\rho_i} \vect{v}_i s_i + n_k
\end{equation}
where $s_i$ is the signal to user $i$ with $\mathbb{E}\{ | s_i |^2 \} = 1$.
The SE at user $k$ is $\log_2(1+\mathrm{SINR}_k)$, where
\begin{equation} \label{eq:SINRkd}
\mathrm{SINR}_k = \frac{ \rho_k | \vect{a}^{\Htran}(\varphi_k,\theta_k) \vect{v}_k |^2  }{ \sum\limits_{i=1, i\neq k}^{K} \rho_i | \vect{a}^{\Htran}(\varphi_k,\theta_k) \vect{v}_i |^2+1}.
\end{equation}
The purpose of this section is to design the precoding; that is, $\{\rho_i\}$ and $\{ \vect{v}_i \}$.
Based on the insights above, we use ZF precoding where user $k$ uses $\vect{v}_k = \vect{w}_k/||\vect{w}_k||$ with
\begin{equation}
\vect{w}_k = \left(\vect{I}_{NM} - \frac{1}{NM}\vect{A}_k (\vect{A}_k^{\Htran} \vect{A}_k )^{-1} \vect{A}_k^{\Htran}  \right)  \vect{a}(\varphi_k,\theta_k)
\end{equation}
where the columns of $\vect{A}_k \in \mathbb{C}^{MN \times K-1}$ are $\vect{a}(\varphi_i,\theta_i)$ for $i=1,\ldots,k-1,k+1,\ldots,K$.
The SINR for user $k$ becomes
\begin{equation} \label{eq:SINRkd2}
\mathrm{SINR}_k =  \mathrm{SNR}_k b_k 
\end{equation}
which depends on $\mathrm{SNR}_k = \rho_kMN$ (i.e., the received SNR of user $k$ without interference) and a constant $b_k\geq 0$ given by 
\begin{equation}
b_k = 1 - \frac{1}{NM}\vect{a}^{\Htran}(\varphi_k,\theta_k)  \vect{A}_k (\vect{A}_k^{\Htran} \vect{A}_k )^{-1} \vect{A}_k^{\Htran} \vect{a}(\varphi_k,\theta_k) .
\end{equation}
It remains to jointly optimize the SINRs by selecting the transmit powers under a total transmit power $Q$. From \eqref{eq:rho_i}, we have that 
\begin{equation} \label{eq:power-constraint}
\sum_{i=1}^{K} q_i = \sum_{i=1}^{K} \rho_i \underbrace{\frac{4\pi r_i^2\sigma^2}{d^2 \cos(\theta_i) \cos(\varphi_i)}}_{= c_i}  \leq Q.
\end{equation}
To determine what is a good power allocation, we define an increasing utility function $U(x)$ and 
 consider the following utility maximization problem:  \vspace{-2mm}
\begin{align} \label{eq:optimization_problem}
\maximize{\rho_1,\ldots,\rho_K} & \quad\sum_{i=1}^{K} U(\rho_i b_i) \\
\,\mathrm{subject \, to} &\quad \sum_{i=1}^{K} \rho_i c_i  \leq Q . \label{eq:power-constraint2}
\end{align}

\begin{lemma}
If $U(x)$ is differentiable and $U'(x) = \frac{d}{dx} U(x)$ is invertible, then the solution to \eqref{eq:optimization_problem} is 
\begin{equation} \label{eq:rhoi_opt}
\rho_i = \frac{1}{b_i} \left[ U'^{-1} \left( \frac{c_i}{\nu b_i }  \right)  \right]_+
\end{equation}
where $[\cdot]_+$ replaces negative values with zero, and the parameter $\nu \geq 0 $ is selected to achieve equality in \eqref{eq:power-constraint2}.
\end{lemma}
\begin{IEEEproof}
Follows from adapting \cite[Th.~3.16]{Bjornson2013d}.
\end{IEEEproof}

If $U(x)=\log(x)$, then we are maximizing the product of the SINRs, which is called proportional fairness \cite{massivemimobook}. We then have $U'(x) = 1/x$ and $U'^{-1} (y) = 1/y$, so that 
\eqref{eq:rhoi_opt} becomes
\begin{equation} \label{eq:rhoi_opt-prodSINR}
\rho_i =  \frac{1}{b_i} \left[ \frac{\nu b_i }{c_i}  \right]_+ = \frac{\nu }{c_i} = \frac{Q }{c_i K} 
\end{equation}
since $\nu = Q/K$ gives equality in the power constraint. This leads to an equal power allocation since $p_i = \rho_i c_i = Q/K$.

If $U(x)=\log_2(1+x)$, then we are instead maximizing the sum SE. It follows that $U'^{-1} (y) = \frac{1}{y \log_e(2)} -1$ and \eqref{eq:rhoi_opt} becomes identical to conventional waterfilling \cite{Tse2005a}:
\begin{equation} \label{eq:rhoi_opt-sumSE}
\rho_i =  \left[ \frac{ \nu }{\log_e(2) c_i} -\frac{1}{b_i}   \right]_+.
\end{equation}

If we want to maximize the harmonic mean of the SINRs, $K/(\sum_{i=1}^{K} \frac{1}{\rho_i b_i})$, we can equivalently set $U(x) = -1/x$. It then follows that $U'^{-1} (y) = \frac{1}{\sqrt{y}}$ and \eqref{eq:rhoi_opt} becomes
\begin{equation} \label{eq:rhoi_opt-prodSINR}
\rho_i =  \frac{1}{b_i} \left[ \sqrt{\frac{\nu b_i }{c_i}}  \right]_+ = \sqrt{\frac{\nu  }{b_i c_i} } = \frac{Q}{\sqrt{b_i c_i}} \Big/ \sum_{k=1}^{K} \sqrt{\frac{c_k}{b_k}}.
\end{equation}

\begin{figure}[t!]
\begin{center} \vspace{-2mm}
	\begin{overpic}[width=\columnwidth,tics=10]{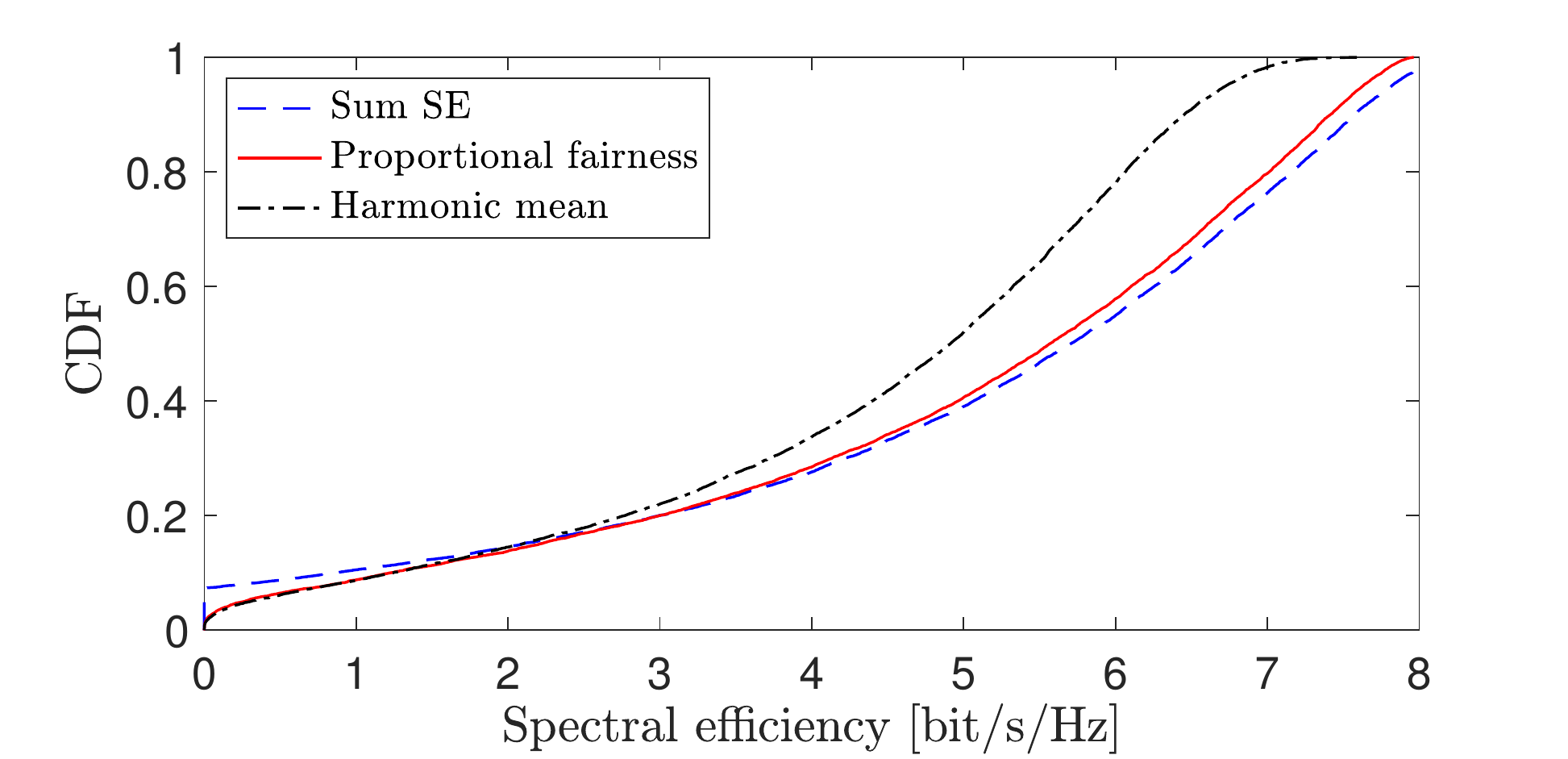}
\end{overpic} 
\end{center} \vspace{-2mm}
\caption{The SE achieved by an arbitrary user when serving $K=5$ with random azimuth angles using ZF precoding.} \vspace{-4mm}\label{simulationUtility}  
\end{figure}

To illustrate the impact of the utility optimization, we consider a surface of size $L=H=10\lambda$ and drop $K=5$ users with uniformly distributed azimuth angles $\varphi_i \in [-\pi/2,\pi/2]$ and $\theta_i=0$. The reference SNR is 0 dB. Fig.~\ref{simulationUtility} shows the CDF of the SE achieved by an arbitrary user when using ZF precoding and the three utilities exemplified above.
Interestingly, the three utilities give similar CDF curves but there are anyway large variations in SE for different user drops. The reason is that the interference is low except when two users happen to get roughly the same angle.

\section{Conclusion}

Spatial interference suppression is important to achieve high spectral efficiency when using an LIS. ZF precoding outperforms MR for practical surface sizes, but we proved that the difference vanishes asymptotically. When using ZF, the power allocation can be efficiently optimized for different utility functions. Although an ideal LIS is a continuous surface, its beam pattern is closely approximated when using discrete antennas of size $\lambda/4 \times \lambda/4$. 
While this paper considered the far-field, the near-field should be analyzed in future work.

\enlargethispage{4mm}

\bibliographystyle{IEEEtran}
\bibliography{IEEEabrv,refs}

\end{document}